# GOTaxon: Representing the evolution of biological functions in the Gene Ontology

Haiming Tang[1,*]; Christopher J Mungall[2]; Huaiyu Mi[1]; Paul D Thomas[1,3,*]

1 Department of Preventive Medicine, University of Southern California, Los Angeles, California, USA

2 Lawrence Berkeley National Laboratory, Berkeley, USA

3 Department of Biological Sciences, University of Southern California, Los Angeles, California, USA

*Corresponding author

Abstract:

The Gene Ontology aims to define the universe of functions known for gene products, at the molecular, cellular and organism levels. While the ontology is designed to cover all aspects of biology in a "species independent manner," the fact remains that many if not most biological functions are restricted in their taxonomic range. This is simply because functions evolve, i.e. like other biological characteristics they are gained and lost over evolutionary time. Here we introduce a general method of representing the evolutionary gain and loss of biological functions within the Gene Ontology. We then apply a variety of techniques, including manual curation, logical reasoning over the ontology structure, and previously published "taxon constraints" to assign evolutionary gain and loss events to the majority of terms in the GO. These gain and loss events now almost triple the number of terms with taxon constraints, and currently cover a total of 76% of GO terms, including 40% of molecular function terms, 78% of cellular component terms, and 89% of biological process terms.

Database URL: GOTaxon is freely available at https://github.com/haimingt/GOTaxonConstraint

## Introduction

The Gene Ontology (GO) project (Ashburner, et al. 2000; Gene Ontology Consortium 2015) is a major bioinformatics initiative to develop a computational representation of our evolving knowledge of how genes encode biological functions. The project has developed formal ontologies that represent over 40,000 biological concepts, and are constantly being revised to reflect new discoveries. The controlled vocabularies of defined terms representing gene product properties that cover three domains: Cellular Component, the parts of a cell or its extracellular environment; Molecular Function, the elemental activities of a gene product at the molecular level, such as binding or catalysis; and Biological Process, operations or sets of molecular events with a defined beginning and end, pertinent to the functioning of integrated living units: cells, tissues, organs, and organisms.

Annotation is the process of assigning GO terms to gene products based on scientific evidence about gene function. Currently, the GO Annotation database (GOA) contains over 280 million annotations for about 4 million different taxonomic groups (April 2017) (Huntley, et al. 2015). Many of these GO annotations have been generated through manual curation from published literatures, others through reviewed computational predictions and minimally supervised automatic prediction pipelines. Although model organisms usually have a large amount of experimental annotations, for many other non-model organisms, experimental annotations are generally lacking and gene annotations of these organisms relay heavily on automated annotation tools. There are a lot of such tools, like InterPro2GO (Burge, et al. 2012) which predicts GO annotations based on predicted domains to which GO classes have been linked and Blast2GO (Conesa, et al. 2005) which is based on sequence similarities between input sequence and pre-annotated genes.

The GO ontology is structured as a directed acyclic graph where each term has defined relationships to one or more other terms in the same domain, and sometimes to other domains. The GO vocabulary is designed to be species-agnostic to enable annotations in all species across prokaryotes and eukaryotes, and single and multicellular organisms. However not all GO classes are observed in all species (Kuśnierczyk, 2008). For example, GO term “heart development” should only be annotated to genes of species that have a “heart”. GO terms related to nucleus and other organelles like “nuclear membrane biogenesis” and “nuclear outer membrane” should not be annotated to genes of prokaryotes, as these cells have no nuclei. These taxonomic constrictions may seem trivial to human curators who have extensive biological training, they are not apparent to automated annotation tools. Thus, taxonomic constraints for GO terms are essential restrictions for automated annotation tools to make correct annotations and avoid “common sense” mistakes.

Back in 2010, the Gene Ontology Consortium formalized an initial list of taxon constraints for GO terms mainly through manual curation (Deegan nee Clark, et al. 2010). Basically, GO taxon constraints set 2 types of taxonomic restrictions for GO terms: “only_in_taxon” and “never_in_taxon” which indicate the GO term should only or never be used to annotate genes of species in the specific taxon. For example, GO:0000330 “plant-type vacuole lumen” is “only_in_taxon” NCBITaxon:33090 “Viridiplantae”, meaning it should only be annotated to genes from species of clade “Viridiplantae”. GO:0000795 “synaptonemal complex” is “never_in_taxon” NCBITaxon:4896 “Schizosaccharomyces pombe”, which prevents the annotation of this GO term to genes from this taxon. In other words, this system described taxon constraints in terms of “annotation white lists” (taxa to which an annotation would be allowed; only_in) and “annotation black lists” (taxa to which an annotation would not be allowed; never_in).

However, the GO taxon constraints list constructed from manual curation in 2010, and extended since that time, is very incomplete. There are only 599 taxon constraint statements in the list, and cover 27.68% of the Gene Ontology terms after propagation to child terms. To address this issue in 2016, Marco Falda et al developed an automatic tool FunTaxIS to add further “only in taxon” or “never in taxon” constraints (Falda, et al. 2016). Unlike manual curation, they chose several “general taxonomic clades” (including Bacteria, Fungi, Viridiplante, Mammalia, Insecta, and a few others) and exploited existing GO annotation frequencies to decide whether each clade should be allowed or forbidden for specific GO terms. FunTaxIS has generated 11.4 million taxon constraints, which cover 94.43% of all GO terms. However, FunTaxIS has a prominent drawback: the taxon constraints are based on the corpus of experimental annotations in the GO knowledgebase, i.e. the experimental observation of gene function as reported in the scientific literature and extracted through the GO biocuration process (Balakrishnan, et al. 2013). They are therefore affected by incompleteness and bias of experimental annotations. For example, if a GO term is only annotated to certain species, it doesn’t necessarily mean this GO term should be constrained to these species or the nearest taxonomic taxa. GO and annotations are subject to the “open world assumption” in which absence of evidence is not considered to be evidence of absence. Thus a function cannot be assumed to be only present in the clade in which it has been observed. Similarly, it cannot be assumed not to be present in a clade in which it has never been observed. “Never_in_taxon” should mean a GO term should not be used for the taxon under all circumstances.

The “only_in_taxon” above provides constraint information that certain GO terms are allowed to be used for annotations of certain species. We suggest that the proper way to represent such constraints from the perspective of evolution, defining the time(s) of emergence (“gain”) and loss of a function in evolutionary history. For example, to express the constraint that a gene can function in the “nucleus” only in eukaryotes, we would express this as “the nucleus emerged on a particular branch of the tree of life, namely the one following LUCA (the last universal common ancestor) and the last common ancestor of all eukaryotes.” We express a “never_in_taxon” type of constraint in terms of when a function is lost along a particular branch of the tree of life. To construct such constraints, the ideal approach would be to have expert biologists manually review the literature on all functions represented by GO terms. However, it would be a lengthy and costly task to manually define taxonomic constraints for a total of more than 42,000 GO terms. We therefore made a first pass at a comprehensive evolution-based constraint assignment by an iterative process where each cycle consisted of five steps: (1) selective manual curation, (2) adding existing constraints from GO as well as other ontologies which are used to define GO terms, (3) logical inference over the ontology structure (propagation of constraints from more general terms to their more specific child terms), (4) combination and integration of constraints collected from various sources and (5) assessment of constraints relative to experimental GO annotations. We first manually curate a seed list of taxon constraints for selected GO terms based on biological knowledge. Second, we combine the manually curated constraints with existing constraints of other sources in GO-plus, and propagate to related GO terms based on the “true path rule”. The constraints are then compared with experimental annotations to find conflicts. We manually check the GO terms with conflicts and those without taxon constraints. Then we modify the seed list and begin another round of propagation. This cycle is repeated until satisfactory.

# Materials and Methods

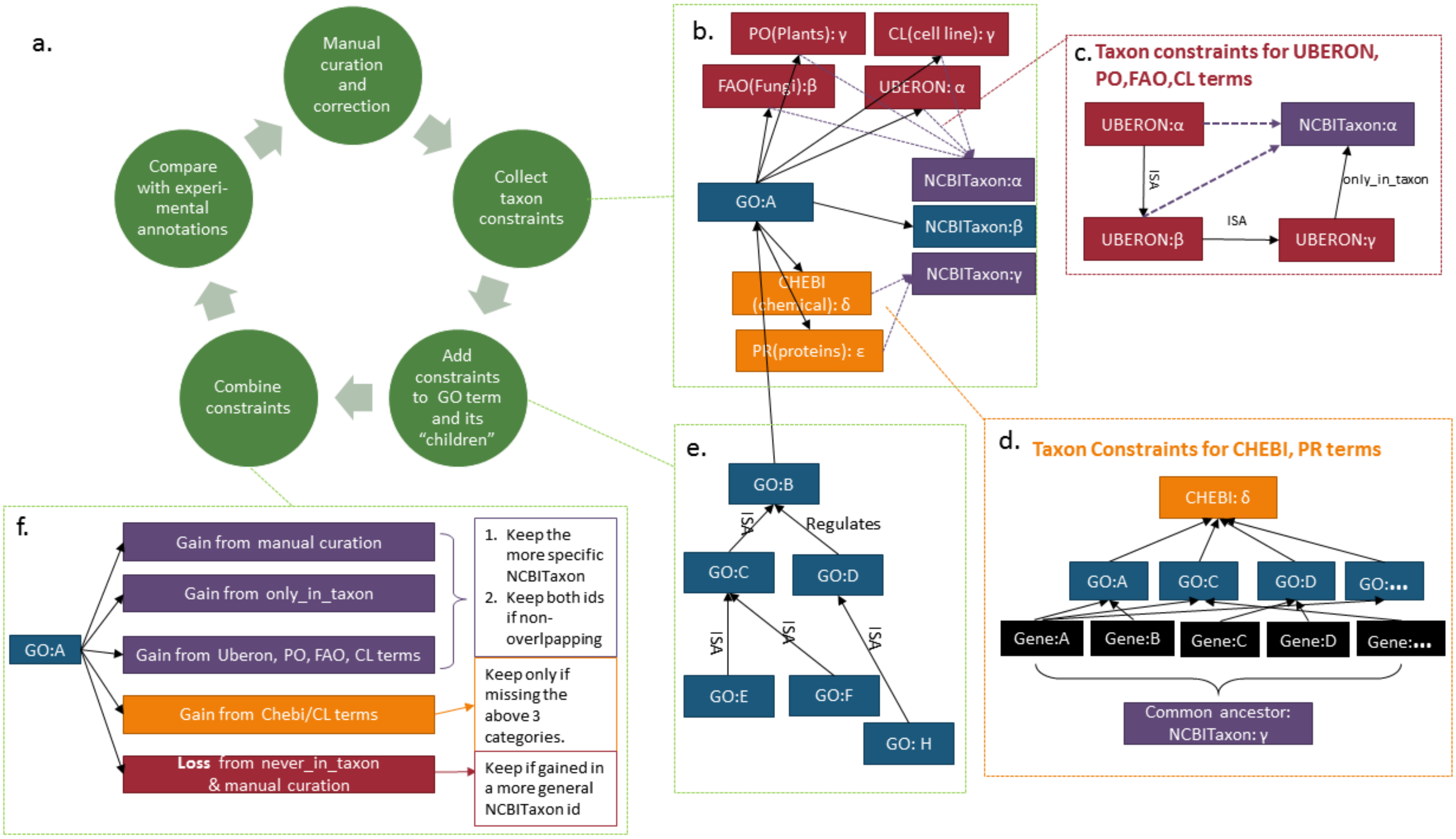

*Figure 1 Workflow of a comprehensive evolution-based constraint assignment for GO terms.*

***a.*** *An iterative process where each cycle consisted of five steps: (1) selective manual curation, (2) adding existing constraints from various sources (3) logical inference over the ontology structure, (4) assignment of constraints by combining constraints from various sources and (5) assessment of constraints relative to experimental GO annotations.* ***b.*** *Collect taxon constraints from various sources. GO:A is a sample GO term which is linked to a taxon constraint: NCBITaxon:β from manual curation seed list or “only_in_taxon”. The other black lines and arrows show references the GO term (GO:A) to other ontology terms which we divide into 2 categories: the more reliable sources: PO (Plants), CL (Cell line), FAO (Fungi) and UBERON (Animals) (red boxes above) and the less reliable sources: ChEBI (Cheminal) and PR (Proteins) (orange boxes below). The dashed line and arrow shows the connection of the other ontology terms with taxon constraints (purple boxes). Finally, GO:A collects constraints NCBITaxon:α, β, γ from all these source.* ***c.*** *An exemplary inference of taxon constraints for UBERON, PO, FAO and CL terms from existing connections with taxon constraints. UBERON:γ has “only_in_taxon” with NCBITaxon:α (solid black line), UBERON:β has a transitive “is_a” relationship with UBERON:γ, and UBERON:α has a trainsitive “is_a” relationship with UBERON:α. Thus we infer UBERON:α and UBERON:β would also be constrained to NCBITaxon:α (purple dashed lines).* ***d.*** *pipeline to infer taxon constraints for ChEBI and PR terms. ChEBI:δ is an exemplary chemical compound which is referenced to many GO terms (blue boxes in the second row), and these GO terms are annotated with many different genes (black boxes in the third row). We find the species of the annotated genes, and then find their common ancestor via trace-up of “is_a” relationship in the NCBITaxon Ontologies. e.*

*Propagation of taxon constraints to related GO terms. This subfigure shows the relation structure of several GO terms. The "ia_a" relationship is transitive, which indicates all children GO terms could share the taxon constraints of a "parent" GO term. Thus GO:E and GO:F could inherit the taxon constraints of GO:C, and GO:C inherits from GO:B and GO:B from GO:A. All other relationships are not considered transitive. In this example, GO:D "regulates" GO:B, and GO:H "is_a" GO:D. Then GO:D doesn't share the taxon constraints of GO:B.* ***f.*** *Assignment of constraints by combining constraints from various sources. Sources are divided into 3 groups where constraints are treated in different ways. The purple blocks are more reliable sources where most specific term is usually chosen. The orange block is the less reliable source where the least specific term is preferred. The red block shows the rules for "loss_at" constraints. For details, please refer to methods part "rules to combine and collapse taxon constraints from different sources".*

## GO-plus information

GO-plus is the fully axiomatised of the GO. It includes cross ontology relationships (axioms) and imports additional required ontologies including ChEBI, cell ontology and Uberon etc. It also includes a complete set of relationships types that link terms within the same ontology or terms from different ontologies (Deegan nee Clark, et al. 2010). Details for different ontologies are summarized in Table1. (http://www.geneontology.org/ontology/extensions/go-plus/)

## GO experimental annotations

The experimental annotations from all species are downloaded from http://geneontology.org/page/download-annotations (Dec 2016). Only annotations with evidence codes: EXP, IMP, IDA, IPI, IGI, IEP are used. As GO terms have a "true-path" hierarchical structure, experimental annotation to a GO term indicates annotation to its all child terms. We find child terms via the "is_a" and "part_of" relationship in GO ontology, and add the experimental annotations to a given term, to all its descendant terms in GO.

## Manual curation and correction

We generated a seed list of taxon constraints based on biological knowledge. This seed list is then used for later propagations. The initial list was neither complete nor (necessarily) correct, but was targeted toward terms toward the root of the GO graph, in order to generate constraints that as parsimonious as possible, and to propagate to a large number of descendant terms in GO. After each propagation cycle, we manually checked the GO terms that are not assigned with taxon constraints as well as taxon constraints that conflict with experimental evidences, modify the list accordingly, and propagate again. The final version of this list is described in detail in the Results section below.

## Collect taxon constraints

In GO-plus, terms of different ontologies form a large network of connections. GO terms can be linked to all ontology types in Table 1. Specifically, as described above, some GO terms are linked to NCBITaxon terms via "only_in_taxon" and "never_in_taxon" relationships. In addition, many GO terms are defined by reference to other ontologies. These include ontologies that describe the anatomical structures of multicellular organisms, including PO (plant anatomy),

FAO (fungal anatomy), UBERON (animal anatomy) ontologies; as well as ontologies that describe cellular and molecular entities, namely CL (cell types in animals), PR (proteins), ChEBI (chemicals), GOCHE (GO CHEmicals). If these ontologies have taxon constraints, then GO terms linked with these ontology terms would inherit the same taxon constraints. The process is illustrated in Figure 1.b. For example, GO:0001525(angiogenesis) is defined in terms of UBERON:0001981(blood vessel). Thus, GO:0001525 would inherit the taxon constraints for "blood vessel": NCBITaxon:7742 (Vertebrata <Metazoa>).

*Table 1 Summary of ontologies in GO-plus*

| Ontologies in GO-plus | Description | number of terms | # terms with taxon constraints | Ref |
|---|---|---|---|---|
| 'GO' | Gene Ontology | 45590 | 599 | (Ashburner, et al. 2000; Gene Ontology Consortium 2015) |
| 'ChEBI' | Chemical Entities of Biological Interest | 6292 | 0 | (Hastings, et al. 2013) |
| 'UBERON' | Cross-species anatomy ontology | 4630 | 204 | (Mungall, et al. 2012) |
| 'CL' | Cell types in animals | 786 | 0 | (Sarntivijai, et al. 2014) |
| 'PR' | protein ontology | 303 | 0 | (Natale, et al. 2014) |
| 'PO' | Plant ontology | 262 | 0 | (Jaiswal, et al. 2005) |
| 'NCBITaxon' | Selected NCBI Taxons | 1330 | | (Deegan nee Clark, et al. 2010; Sayers, et al. 2009) |
| 'OBA' | Ontology of Biological atrributes | 113 | 0 | |
| 'GOCHE' | GO CHEmicals | 77 | 0 | (Hill, et al. 2013) |
| 'SO' | sequence ontology | 37 | 0 | |
| 'FAO' | Fungi anatomy | 29 | 0 | |
| 'PATO' | phenotypic qualities (properties, attributes or characteristics) | 10 | 0 | |

As PO(plants), FAO(Fungi), UBERON(Animals), and CL (animal cell lines) ontology terms also include the "only_in_taxon" constraints, we process them in a similarly way describe above. We first extract the set of terms that are linked to NCBITaxon, like UBERON:γ and NCBITaxon:α illustrated in Figure 1.c. Then we add the same taxon constraints to all children of this set of terms, like UBERON:α and UBERON:β in Figure 1.c. CL (Animal cell lines) are usually linked to UBERON (Animals) via "part_of" relationship. Thus CL terms inherit the taxon constraints from the UBERON terms. Besides, CL terms could also inherit taxon constraints from ancestor (more general) CL terms.

Due to the complexity of the relation network, there are scenarios when a term can get constraints from multiple sources. In this case, we assign the innermost (the most specific or youngest) NCBITaxon id via the NCBITaxon ontology network.

After combination and integration processes above, some terms may still have no taxon constraints. Then, we assign these terms with constraints by the natural properties of the terms. Specifically, the PO (plants) terms are assigned with NCBITaxon: 33090 (Viridiplantae), FAO(fungi terms) are assigned with NCBITaxon:4751 (Fungi), UBERON(animal structure) are assigned with NCBITaxon: 33213 (Bilateria), CL (cell lines) are assigned with NCBITaxon: 33213 (Bilateria) except a few high level terms near the root of the ontology hierarchies pinpointed by manual examinations. Examples include "UBERON:000000 process entity" and "UBERON:0000061 anatomical structure" etc. The full list of taxon constraints for PO(plants), FAO(Fungi), UBERON(Animals), CL (animal cell lines) ontologies are in Supplemental file 1.

PR(proteins) and ChEBI (chemicals) do not have taxon constraints already associated with them in GO-plus, thus we use a method similar in spirit to FunTaxIS to assign taxon constraints for these terms. We aggregate experimental evidence over all GO terms that reference each protein or chemical even if these GO terms are not connected in the GO hierarchical structure, we then collect the genes that are annotated with these GO terms, and find the common ancestor of the species associated with these genes. Finally, we assign the common ancestor as the taxon constraint for the PR or ChEBI terms, This pipeline is illustrated in Figure 1.d. The full list of taxon constraints for ChEBI and PR ontologies are in Supplemental file 2.

## Find all propagatable children of GO terms

In the Gene Onotolgy network, GO terms are connected with other GO terms with several different types of relationships. Aside from the "is_a" relationship that defines the "true-path" in GO hierarchy, there are about 70 additional relationships like "regulates", "part_of", "results_in", "capable_of" etc. The complete list of relationships and the total number of each relationship are in Supplemental file 3. To fully utilize these relationships, we extend "true child terms" as defined by the "true-path" in GO hierarchy to a concept we call "propagatable child terms": the set of "child" GO terms which can inherit the taxon constraints from a "parent" GO term. These "propagatable child terms" include all "true child terms" which are connected via transitive "is_a" relationship and all GO terms that are directly linked to the "parent" GO term via any type of relationships. The example in figure1.e shows that GO:C and GO:D “is_a” GO:B; GO:D “regulates” GO:B; GO:E and GO:F “is_a” GO:C; GO:H “is_a” GO:D. Thus GO:E, GO:F and GO:C are “true child terms” of GO:B via the transitive “is_a” relationship, and they share the constraints of GO:B. WGO:D is a “propagatable child term” of GO:B, which also shares the constraints of GO:B. GO:H is a “true child” of GO:D via “is_a” relationship, but GO:H is not a propagatable child of GO:B, as the “regulates” relationship is not transitive.

## Rules to combine and collapse taxon constraints from different sources

We can build taxon constraints for GO terms from different sources: the relationships between GO terms with NCBITaxon in our seed list, the relationships between GO and NCBITaxon in go-plus, the relationships between GO and PO (plants), FAO (Fungi), UBERON (Anatomy) and CL (cell lines), the relationships between GO and ChEBI (chemical) and PR (proteins), and the “child-parent” relationship between GO terms.

In practice, we often get enormous amount of constraints for each GO term and we need to integrate these taxon constraints for conclusive “Gain at” taxon constraint. The rules are illustrated in figure 1.f. There are four main sources for “Gain at” taxon constraints: seed list of manual curations, “only_in_taxon” in GO-plus, related PO, FAO, UBERON, CL terms that have taxon constraints, and related ChEBI and PR terms with taxon constraints. For the first three sources, the constraints are from manual curation. Therefore, these taxon constraints are generally reliable, and the constraints from different sources should not conflict with each other. Thus we assign the most recent taxon constraint to the GO term, as it is consistent with the older taxon constraints from other sources. For example, we assign “NCBITaxon:7742(Vertebrata)” to term 'GO:0001503' (ossification) as this term inherits three constraints from its ancestors: “NCBITaxon:33213(Bilateria)”, “NCBITaxon:2759(Eukaryota)” and “NCBITaxon:7742(Vertebrata)”, we choose Vertebrata as the taxon constraints for the GO term as Vertebrata is the youngest taxon, and it also satisfies the constraint of both Bilateria and Eukaryota. If there exist two or more taxa that do not overlap with each other among the collections, then we assign both taxa. For example: “gain_at” NCBITaxon:33090(Viridiplantae) or “gain_at” NCBITaxon:6072 (Eumetazoa)” are assigned to term GO:0000803 ‘sex chromosome’.

As constraints of ChEBI and PR terms are from assembled experimental evidence in GO annotations (and therefore subject to the open world assumption), if the experimental evidences are sparse, the taxon constraints may not be complete. Thus we handle this source with extra attention. We assign the taxon constraint from these sources only if the term is not assigned a constraint using any of the more reliable methods described above. Furthermore, in contrast to other constraints, if a GO term (including its descendants) can be linked to multiple ChEBI /PR terms, we assign the *least* specific (oldest) taxon.

The sources of taxon constraints for “evolutionary loss” are always manually curated. These come from the manually curated seed list and manually assigned “never_in_taxon” constraints in GO-plus. Any functional characteristic must have been present in an ancestor in evolutionary history, and then after that the characteristic could get lost in one of the ancestor's descendants. Therefore, we assign a "loss at a given taxonomic clade" for a GO term only if it is simultaneously assigned with "gain at" a more ancient taxonomic clade.

### Comparison of the finalized GO taxon constraints with experimental evidences

From experimental annotations, we collect all genes that are annotated to a GO term, and then find the species for these genes. So we get a list of species for each GO term with experimental annotations. Next, we compare the list of species with constraint taxon id of the GO term. A conflict is found if any species in the list is not a descendant of a "gain_at" taxon id or the species is a descendant of a "loss_at" taxon id. We then manually review these conflicts. These situations can indicate an error in the constraints manually assigned to either a GO term or to PO, FAO, UBERON or CL, or in the constraints inferred from annotations to GO terms that use ChEBI or PR terms. We then correct any found errors and rerun the whole propagation process. We also found cases where the conflicts came from errors in experimental GO annotations. We reported these cases in Supplementary File 5, and have reported to the GO Consortium. The errors are already being corrected in the GO knowledgebase.

## Results

Here we construct a comprehensive set of taxon constraints for GO terms using an iterative process that combines manual curation and automated propagation from "parent" GO terms to "child" terms. The GO taxon constraints are freely available at https://github.com/haimingt/GOTaxonConstraint . The taxon constraints are encoded by "gain_at" and "loss_at" of specific taxon, which represent a clade in the tree of life (properly speaking this means the evolutionary event occurred along the branch immediately prior to an ancestor that defines a taxonomic clade). This can be interpreted as dating, relative to speciation events that define common ancestors in the tree of life, the evolutionary emergence of a function in a clade, and, later, its loss in some sub-clade. Thus, a GO term can be applied to a gene from a species within clade in which it was gained, and should not be applied either outside of the clade in which it was gained, or to species inside the clade in which it was subsequently lost. If a GO term was gained along multiple distinct branches in the species tree, it indicates that the function class was gained independently in multiple clades (convergent evolution), thus it could be used to annotate species that belongs to any of these clades. An example of convergent evolution is "multicellular organismal process" (GO:0032501), which arose separately in multiple clades including the animals, multicellular plants and multicellular fungi. This class of process was subsequently lost as a possible gene function in multiple fungal lineages, leading to the extant organisms commonly known as yeasts (single-celled fungi).

There is a total of 45589 GO terms in GO-plus, with 24.91% of them are molecular function terms, 9.21% cellular component terms and 65.88% biological process terms. 34654 GO terms (76.01%) are constructed with taxon constraints. Biological process terms have higher percentage of constraint coverage (89.24%) than cellular component (77.23%) and molecular function (40.29%). 9532 GO terms are annotated with taxon constraints from ChEBI terms only and should be considered less reliable; these account for 2204 (46%) of all molecular function GO terms with taxon constraints. 15.78% GO terms are annotated with “Loss” taxon constraints, and the vast majority of them are biological process terms. 25.96% GO terms are annotated with more than one “Gain” taxon constraints, the clear majority of them are also biological process terms, accounting for 31.38% of all biological process terms with taxon constraints.

Detailed statistics are listed in Table 2. We see a significant difference in taxon constraints among the three categories of GO terms. Molecular functions depict the molecular activities of gene products, including binding to things, enzyme activity, molecular transporter activity, and molecular receptor activity. Thus, molecular function terms are often associated with molecular entities that are encoded by ChEBI (chemical) and PR (protein) terms. This explains why nearly half of the taxon constraints of molecular function terms are from ChEBI and PR terms, and most of these constraints are “Gain at root” which means the molecular function term is available to annotated to any species. The biological processes that a gene product is involved in, on the other hand, generally consists of many activities, which involve diverse molecular functions of many genes. This is the reason why larger percentage of biological process terms is annotated with taxon constraints than molecular functions.

*Table 2 Summary of taxon constraints for GO terms*

| | Total number of GO terms | GO terms with taxon constraints | Go terms with taxon constraints from ChEBI terms | Go terms with "Loss" taxon constraints | Go terms with more than 1 "Gain" taxon constraints |
|---|---|---|---|---|---|
| molecular_function | 11355 | 4586 (40.39%) | 2204 (48.06%) | 36 (0.78%) | 168 (3.66%) |
| cellular_component | 4199 | 3264 (77.73%) | 84 (2.57%) | 119 (3.65%) | 417 (12.78%) |
| biological_process | 30035 | 26804 (89.24%) | 7244 (27.03%) | 5314 (19.83%) | 8412 (31.38%) |
| Total | 45589 | 34654 (76.01%) | 9532 (27.51%) | 5469 (15.78%) | 8997 (25.96%) |

## Most widely annotated taxa for GO taxon constraints

In Table3, we summarize the 10 most widely used taxa for GO taxon constraints construction, and collapse the others to “other taxon”.

We see NCBITaxon:1 root (last universal common ancestor) is the most widely used taxon, which implies these terms could be used for annotations of genes from all species of cellular organisms. The taxon constraint categories generally represent lineages that underwent relatively large evolutionary changes, such as the evolution of the eukaryotic cell (Taxon Eukaryota used in 4656 terms, 6.91% of all GO terms constraints), multicellular organisms (Taxon Eumetazoa/Bilateria, Dikarya 8.86%, Embryophyta 11.01%) or multicellular life stages (Taxon Dictyostelium used in 8.27%). The largest number of losses occurs at Saccharomycetales and Schizosaccharomyces pombe due to the reversion to a unicellular lifestyle from multicellular ancestors (15.03% of all constraints and 86.61% of all losses). The final manual curations for seed list

In Table 4, we summarize all manually curated taxon constraints, and the total number of GO terms that have taxon constraints. As some taxon constraints are applied to many GO terms, only some GO terms are listed in the “example” column to save space. The full list is available at Supplemental file 4.

We have manually curated a total of 197 GO terms. 18 terms have constraints “gain_at” root. These terms including cellular components that are found in all cellular organisms like chromosome and cytoskeleton, and some high-level GO terms like cytokinesis, behavior and cell adhesion.4 terms have constraints “gain_at” root and “loss_at” Eukaryota (indicating constraints to only prokaryotes and archaea), these GO terms include FtsZ and Cdv dependent cytokinesis, cytoplasmic chromosome and transcription antitermination; these GO terms are assumed to have been present in LUCA, but lost in Eukaryotes. 57 terms have constraint “gain_at” Eukaryota, including eukaryotic organelles like nucleus, mitochondrion, lysosome and some molecular functions like MAPK cascade, G-protein coupled receptor activity and biological processes like gamete generation and fertilization. 4 terms for bacterial specific cellular component are represented as “gain_at” Bacteria; 2 Archaeal cellular component specific terms are annotated as “gain_at” Archaea. “Cell envelope” is found in both Bacteria and Archaea, and is represented as “gain_at” Archaea and Bacteria, as these are distinct types of envelopes that likely resulted in similar characteristics through convergent evolution. Similarly, “cell wall” is represented as convergent evolution in four lineages: Archaea, Bacteria, Fungi, and Viridiplantae.

*Table 3 Summary of Taxon that are most widely used for taxon constraints of GO terms*

| Taxon id and name | | biological process | molecular function | cellular component | Total | Gain/Loss Total |
|---|---|---|---|---|---|---|
| NCBITaxon:1(root) | Gain | 12862 | 996 | 3778 | 17636 | 17636 (26.18%) |
| | Loss | 0 | 0 | 0 | 0 | |
| NCBITaxon:3193 (Embryophyta) | Gain | 7232 | 125 | 62 | 7419 | 7419 (11.01%) |
| | Loss | 0 | 0 | 0 | 0 | |
| NCBITaxon:4892 (Saccharomycetales) | Gain | 10 | 0 | 0 | 10 | 5069 (7.53%) |
| | Loss | 4926 | 109 | 24 | 5059 | |
| NCBITaxon:6072 (Eumetazoa) | Gain | 2640 | 235 | 111 | 2986 | 2986 (4.43%) |
| | Loss | 0 | 0 | 0 | 0 | |
| NCBITaxon:451864 (Dikarya) | Gain | 5496 | 128 | 24 | 5648 | 5965 (8.86%) |
| | Loss | 312 | 1 | 4 | 317 | |
| NCBITaxon:2759 (Eukaryota) | Gain | 3049 | 1287 | 312 | 4648 | 4656 (6.91%) |
| | Loss | 3 | 4 | 1 | 8 | |
| NCBITaxon:33213 (Bilateria) | Gain | 3577 | 264 | 126 | 3967 | 3967 (5.89%) |
| | Loss | 0 | 0 | 0 | 0 | |
| NCBITaxon:5782 (Dictyostelium) | Gain | 5437 | 109 | 24 | 5570 | 5570 (8.27%) |
| | Loss | 0 | 0 | 0 | 0 | |
| NCBITaxon:4896 (Schizosaccharomyces pombe) | Gain | 11 | 0 | 0 | 11 | 5077 (7.54%) |
| | Loss | 4928 | 114 | 24 | 5066 | |
| NCBITaxon:7742 (Vertebrata) | Gain | 2932 | 54 | 8 | 2994 | 3005 (4.46%) |
| | Loss | 9 | 2 | 0 | 11 | |
| other taxa | Gain | 3572 | 852 | 359 | 4783 | 6012 (8.92%) |
| | Loss | 1059 | 140 | 30 | 1229 | |
| Total | | 58055 (86.18%) | 4420 (6.56%) | 4887 (7.25%) | 67362 | 67362 |

"Multicelluar organisms" include animals, plants and most fungi except the yeasts. Thus, we constrained "Multicelluar organisms" related terms like "multicellular organismal process" and "cell junction" to "gain_at" Eumetazoa and Viridiplantae and Dikarya, with subsequent "loss_at" Saccharomycetales and Schizosaccharomyces pombe. At the same time, we constrained "Single cellular organism" related terms like "adhesion between unicellular organisms" to "gain_at" LUCA and Saccharomycetales and Schizosaccharomyces pombe, with subsequent (to LUCA) "loss at" Eumetazoa, Viridiplantae or Dikarya.

We constrained "Apoptosis" related terms to "gain _at" Opisthokonta and Embryophyta as the term is used to describe two nonhomologous (convergently evolved) processes.

Term "mating" is constrained to "gain_at" Opisthokonta. Separation of "sex" is observed in animals, plants and fungi, however there is no consensus as to when "sex" first emerged in evolutionary history or if "sex" has evolved separately multiple times between and even within these taxa. Thus, for now we simply correlated "sex" with multicellularity, i.e. "gain_at" Eumetazoa, Viridiplantae and Dikarya, and more specific terms like "sexual sporulation" to Dikaryta, and "femal gamete generation" to Eumetazoa, Embryophyta and Dikarya.

Animals are probably the best-studied organisms on earth (leading to some inevitable bias in the GO) but also represent the emergence of many innovations. GO terms like nervous system development, heart development, muscle organ development and spermatogenesis are constrained to this taxon. More specific terms related with animal structure development and functioning like ectoderm development, sensory perception, neurological system process, muscle contraction is constrained to "Eumetazoa". For similar reasons, vertebrates are also a common place for functions to be "gained_at". The skeletal system and blood first emerged in this clade, as well as the acquired immune system. Thus, terms like "skeletal system development", "angiogenesis", "blood coagulation", "immunoglobulin complex", "complement activation" are constrained to "Vertebrata". Two terms "hemopoiesis" and "inflammatory response" are constrained to "gain_at" vertebrata or "Arthropoda", as these terms are also used to describe analogous processes in arthopods: Hemolymph circulates in the interior of the arthropod body, and arthopods possess an inflammatory response. "Immune system process" is constrained to "gain_at" Eumetazoa and Embryophyta as again these are analogous processes.

Term "mammary gland development" is constrained to Mammalia. "Pollen tube adhesion" and "seed oilbody biogenesis" are constrained to flowering plants: Spermatophyta. Thylakoid is the site of the light-dependent reactions of photosynthesis, it is constrained to "gain_at" Cyanobacteria and Viridiplantae. The term "sorocarp" is the fruiting body of Dictyostelia which are a group of cellular slime molds, or social amoebae, and is constrained to "gain_at" Dictyostelium.

Term “rhabdomere membrane” is a cellular component of compound eyes of arthropods; related GO terms are constrained to gain_at Arthropoda. Rhoptry is a specialized secretory organelle connected by thin necks to the extreme apical pole of parasite, related terms are constrained to “gain_at” Apicomplexa.

## Seed lists from other sources

Apart from 197 manually curated GO terms, there are 599 GO terms whose taxon constraints are from the “only_in_taxon” or “never_in_taxon”constraints in GO-plus, 5593 terms whose taxon constraints are from PO (Plant), FAO (Fungi), UBERON (Anatomy) and CL (Cell line) terms, and 3104 terms whose taxon constraints are from ChEBI (chemical) and PR (protein) terms. The full seed list contains 9296 terms. These taxon constraints are propagated to 34654 terms. Detailed statistics are listed in Table 5.

In Figure 2, we plot a venn diagram to show how the taxon constraints of the GO terms in the “seed list” are propagated to other GO terms. The diagram shows a GO term could inherit constraints from various combinations of the sources. For example, only 1625 GO terms have “gain_at” constraints from all 4 sources, and there are 7666 (6041+1625) GO terms that would inherit taxon constraints from manual curation, only_in_taxon, PO (Plant), FAO (Fungi), UBERON (Anatomy) and CL (Cell line) terms. The most specific taxon out of all these taxon constraints is chosen to annotate these GO terms. For the time being, taxon constraints for ChEBI and PR terms are constructed from assembling experimental evidences, and they are very likely to result in overly stringent constraints due to incompleteness of GO annotations and the open world assumption. To minimize this problem, taxon constraints from these terms are only used when other sources of constraints are all missing. We aggregate all experimental annotations to GO terms that are linked to a given ChEBI term, and assume that the compound was present in the common ancestor of all the organisms for which an annotation was made. This approach will neglect convergent evolution but err on the side of a permissive constraint. As a result, all but 15 of ChEBI-derived constraints are “gain_at” LUCA. We manually checked these terms and found 2 errors that we subsequently corrected by adding manual constraints.

PO (plants) and FAO (Fungi) ontologies have a relatively small number of terms, and most of them are assigned to taxon Viridiplantae and Fungi. Although this approach will miss many lineage-specific fungal structures, it is a large improvement over having no constraint at all. For UBERON and CL ontologies, we automatically assigned “Eumetazoa” taxon to terms that still have no constraints after the construction process for these terms (Details in methods part). This process has yielded some problems: the taxon of UBERON and CL ontologies in-turn propagate “Eumetazoa” taxon to some GO terms, and result in conflicts with experimental evidences (Details below). We have examined the terms that caused conflicts and some high-level terms of both ontologies, and manually corrected the automatically assigned “Eumetazoa” taxon.

*Table 4 manually curated taxon constraints*

<table>
<tr><th>Taxon constraints:</th><th>Total numbe</th><th>Examples:</th></tr>
<tr><td>Gain| root;</td><td>18</td><td>chromosome;cytoskeleton;cell differentiation</td></tr>
<tr><td>Gain| root;>Loss| Eukaryota;</td><td>4</td><td>FtsZ-dependent cytokinesis;Cdv-dependent cytokinesiscytoplasmic chromosome;transcription antitermination</td></tr>
<tr><td>Gain| Eukaryota;</td><td>57</td><td>organelles like nucleus;mitochondrion ； MAPK cascade; gamete generation</td></tr>
<tr><td>Gain| Bacteria;</td><td>4</td><td>bacterial-type flagellum;Gram-negative-bacterium-type cell wall biogenesis;type IV pilus biogenesis;bacterial-type flagellum-dependent cell motility;</td></tr>
<tr><td>Gain| Archaea;</td><td>2</td><td>archaeal-type flagellum;archaeal-type flagellum-dependent cell motility;</td></tr>
<tr><td>Gain| Archaea; Bacteria;</td><td>1</td><td>cell envelope;</td></tr>
<tr><td>Gain| Archaea; Bacteria; Fungi; Viridiplantae;</td><td>2</td><td>cell wall;cell wall biogenesis;</td></tr>
<tr><td>Gain| Bacteria; Dikarya;</td><td>1</td><td>conjugation with cellular fusion ;</td></tr>
<tr><td>Gain| root; Saccharomycetales; Schizosaccharomyces pombe;>Loss| Dikarya; Embryophyta; Eumetazoa;</td><td>1</td><td>adhesion between unicellular organisms;</td></tr>
<tr><td>Gain| root;; Schizosaccharomyces pombe;>Loss| Eumetazoa; Viridiplantae; Dikarya;</td><td>1</td><td>conjugation with mutual genetic exchange;</td></tr>
<tr><td>Gain| Dikarya; Embryophyta; Eumetazoa; Dictyostelium;>Loss| Saccharomycetales; Schizosaccharomyces pombe;</td><td>2</td><td>extracellular matrix;multicellular organismal process;</td></tr>
<tr><td>Gain| Opisthokonta; Embryophyta;</td><td>2</td><td>apoptotic process;negative regulation of apoptotic process;</td></tr>
<tr><td>Gain| Opisthokonta;</td><td>1</td><td>mating;</td></tr>
<tr><td>Gain| Bilateria;</td><td>30</td><td>ectoderm development;sensory perception;neurological system process;muscle contraction</td></tr>
<tr><td>Gain| Eumetazoa; Embryophyta; Dikarya;</td><td>1</td><td>female gamete generation;</td></tr>
<tr><td>Gain| Eumetazoa; Viridiplantae; Dikarya;</td><td>1</td><td>sex determination;</td></tr>
<tr><td>Gain| Eumetazoa; Viridiplantae;</td><td>22</td><td>multi-multicellular organism process;single-multicellular organism process;cell junction</td></tr>
<tr><td>Gain| Eumetazoa;</td><td>11</td><td>nervous system development;heart development;muscle organ development;spermatogenesis</td></tr>
<tr><td>Gain| Vertebrata <Metazoa>;</td><td>20</td><td>skeletal system development;hemopoiesis;inflammatory response</td></tr>
<tr><td>Gain| Vertebrata <Metazoa>; Arthropoda;</td><td>2</td><td>inflammatory response;hemopoiesis;</td></tr>
<tr><td>Gain| Mammalia;</td><td>1</td><td>mammary gland development;</td></tr>
<tr><td>Gain| Spermatophyta;</td><td>2</td><td>pollen tube adhesion;seed oilbody biogenesis;</td></tr>
<tr><td>Gain| Cyanobacteria; Viridiplantae;</td><td>1</td><td>thylakoid;</td></tr>
<tr><td>Gain| Embryophyta; Eumetazoa;</td><td>1</td><td>immune system process;</td></tr>
<tr><td>Gain| Viridiplantae;</td><td>1</td><td>plastid translation;</td></tr>
<tr><td>Gain| Viridiplantae; Apicomplexa;</td><td>1</td><td>plastid;</td></tr>
<tr><td>Gain| Dikarya;</td><td>1</td><td>sexual sporulation;</td></tr>
<tr><td>Gain| Dictyostelium;</td><td>1</td><td>sorocarp development;</td></tr>
<tr><td>Gain| Apicomplexa;</td><td>1</td><td>rhoptry;</td></tr>
<tr><td>Gain| Arthropoda;</td><td>1</td><td>rhabdomere membrane biogenesis;</td></tr>
</table>

*Table 5. Summary of "seed list" from all sources.*

| | GO terms in seed list: Manual curation | GO terms in seed list: only/never_in_taxon | GO terms in seed list: PO/FAO/UBERON/CL | GO terms in seed list: ChEBI/PR | Total GO terms from all sources |
|---|---|---|---|---|---|
| molecular_function | 8 | 30 | 1026 | 268 | 1324 (14.24%) |
| cellular_component | 53 | 197 | 7 | 88 | 317 (3.41%) |
| biological_process | 136 | 372 | 4560 | 2748 | 7655 (82.35%) |
| Total | 197 (2.12%) | 599 (6.44%) | 5593 (60.17%) | 3104 (33.39%) | 9296 (26.82%) |

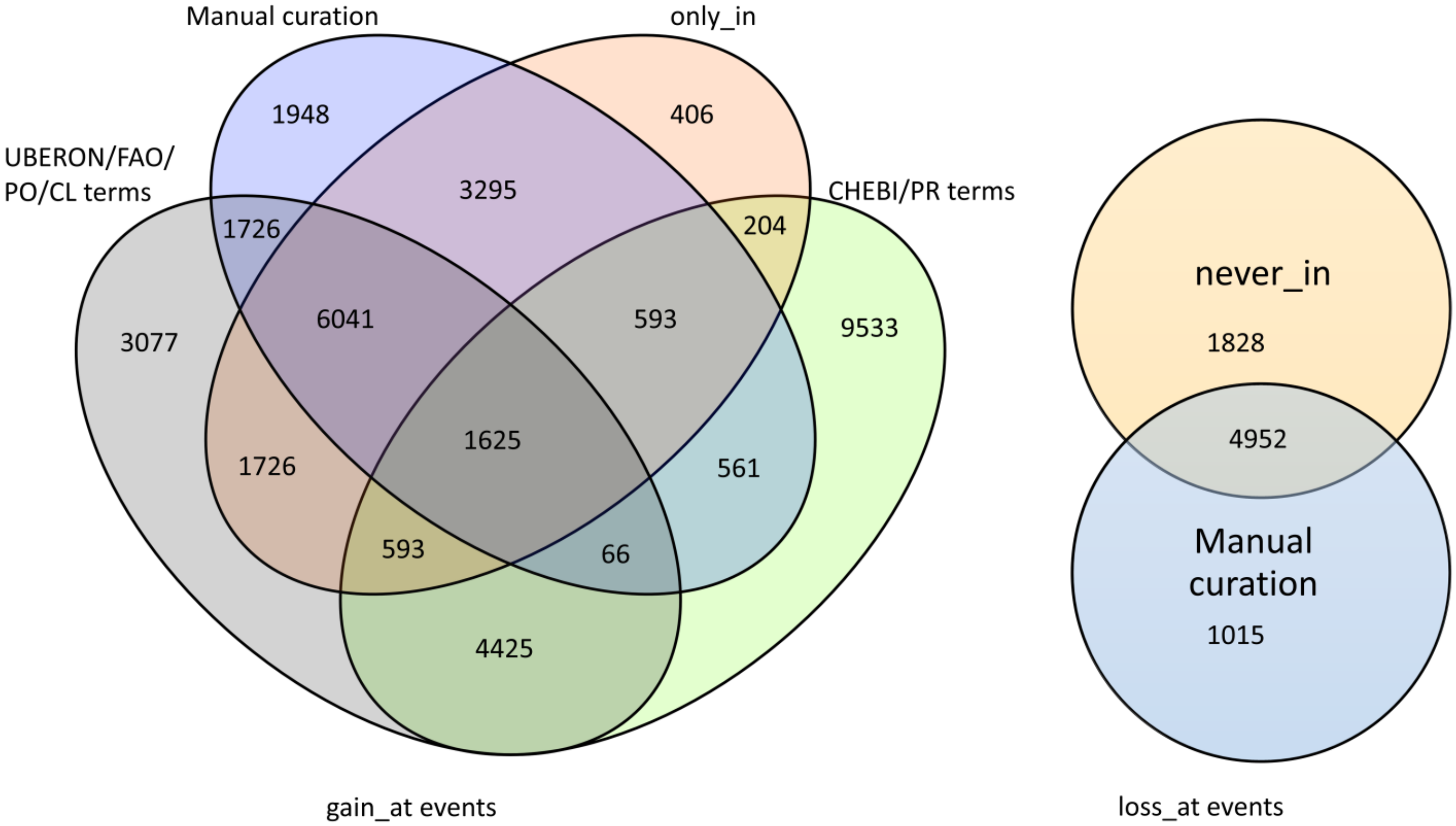

*Figure 2. GO terms propagated from "seed list" of different sources*

*Figure 2 shows how the taxon constraints of the GO terms in the "seed list" are propagated to other GO terms. The 2 venn diagrams show gain_at events and loss_at events separately. The diagram shows a GO term could inherit constraints from various combinations of the sources. For example, there are 7666 (6041+1625) GO terms that would inherit "gain_at" taxon constraints from manual curation, only_in_taxon, PO (Plant), FAO (Fungi), UBERON (Anatomy) and CL (Cell line) terms. And 4952 GO terms that would inherit "Loss at" taxon constraints from both "never_in_taxon" and manual curation.*

*Table 6. Comparision of constructed taxon constraints with experimental evidences and FunTaxIS results*

| | | # conflicts | # BP terms | # MF terms | #CC terms | total GO terms | Most common species: and # occurrence | second most common species and # occurrence |
|---|---|---|---|---|---|---|---|---|
| **Conflict with experimental evidences** | species with experimental evidences is outside of Gain taxon | 267 | 195 | 5 | 22 | 222 | Bacteria 84 | Terrabacteria group 50 |
| | species with experimental evidences is within Loss taxon | 0 | 0 | 0 | 0 | 0 | | |
| **Conflict with Funtaxis** | "only_in" taxa of FunTaxIS is outside of Gain taxon | 3313 | 1090 | 24 | 152 | 1266 | Bacteria 362 | Terrabacteria group 295 |
| | "only_in" taxa of FunTaxIS is within Loss taxon | 0 | 0 | 0 | 0 | 0 | | |
| | "never_in" taxa of FunTaxIS is in Gain taxon | 705645 | 25922 | 3619 | 2784 | 32325 | saccharomyceta 22701 | Embryophyta 20616 |
| | "never_in" taxa of FunTaxIS is mom of Gain taxon | 80347 | 10083 | 240 | 420 | 10743 | Bacteria 20442 | Opisthokonta 7722 |

## Improvement over previous version of GO taxon constraints

Gene Ontology Consortium first formalized an initial list of taxon constraints for GO terms in 2010 (Deegan nee Clark, et al. 2010). The taxon constraints are composed of 2 relationships: "only_in_taxon" and "never_in_taxon" where a GO term and its subtypes and parts should only or never be used for annotation of genes from species of the specific taxon. The "only_in_taxon" also indicates that the GO term should never be used outside of the named taxonomic group. For function terms that have evolved in multiple taxa independently, "only_in_taxon" could hardly define the correct taxonomic constraints. For example, "cell wall" may have evolved independently in Archaea, Bacteria, Fungi and Green plants, but it would not be straightforward to define the taxonomic constraint for this term as “only_in_taxon” LUCA and add “never_in_taxon” to all Eukaryotes except Fungi and Green plants. To solve this problem, we proposed a better way to represent taxonomic constraints from the perspective of evolution, using “gain_at” and “loss_at” events, which indicate the times(s) of emergence (“gain”) and loss of functions in evolutionary history. For the example above, the taxonomic constraints of “cell wall” is “gain_at Archaea;Bacteria;Fungi;Viridiplantae”. In total, there are around 9000 GO terms, which have more than 1 “Gain” taxon constraints.

In the previous version of GO taxon constraints, there are only 599 constraint statements constructed from mainly manual curation. Even after propagation to subtypes and parts, these constraints only over 27.68% of the Gene Ontology terms. This update has greatly expanded the taxon constraints construction using a semi-automatic pipeline to summarize taxon constraints from different sources. The final taxon constraint statements cover 76.01% of all terms. Besides, most of the GO terms that have no taxon constraints are molecular function terms, which are applicable to species of all taxa.

## Comparison of our construction of GO taxon constraints with experimental evidences

We use gene annotations with experimental evidence codes to find conflicts and errors in our taxon constraint construction process. There are two types of such conflicts, the first type is species with experimental evidences is outside of Gain taxon and the other type is species with experimental evidences is within the Loss taxon. No conflicts of the second type are observed. The conflicts results are summarized in Table 6. A complete list of conflicts is in supplemental file 5.

If a species with experimental evidences is outside of Gain taxon, it is possible that the taxon constraints of our construction is too strict, and the common ancestor of the constrained taxon and the species should be used instead. In early iterations of our process we identified many such errors, which were corrected in later iterations. One example is in our initial list, we constrained “behavior” to animals. However, we found that the term has been used much more generally in GO annotation (particularly in plants), and we changed the constraint to LUCA. Another example is the GO term “muscle fiber development”, which was initially constrained to “gain_at” chordata, but has experimental evidence from insects. It could have got the correct constraint of “gain_at” Bilateria from our manual curation list, however it also gets the constraint from a CL term “striated muscle cell” which is incorrectly constrained to “Chordata”. The taxon Chordata is then choosen at the propagation process as it is more specific than Bilatera. Thus, we corrected the constraint for the CL term. Another possibility of this type of conflict comes from wrong experimental annotations. After numerous rounds of manual revision, there are still 267 conflicts involving 222 GO terms (see Supplemental File 5). We believe that most of these conflicts are from errors in experimental annotations, and we have submitted this list to the GO Consortium for review. Examples, include “barrier septum assembly,” which has experimental evidence from bacteria, “ovarian follicle development” with experimental evidence from fly, “activation of immune system” has experimental evidence from Plasomidum, “regulation of apoptotic process” has experimental evidence from Bacteria and so on. Notably, a recurrent theme is that immune responses related GO terms are annotated to genes of infectious microbes like some bacteria and eukaryotic parasites. These organisms do not possess immune systems, thus their genes cannot be said to function in their own immune response (which is what the annotation actually means). However, it is clear that the curators meant to associate the gene with its effect on the host organism’s immune system, and more recently added GO terms for symbiotic interactions would be more appropriate.

## Conflicts with FunTaxIS

FunTaxIS constraints are coded as “in taxon” or “never_in_taxon” using the frequency of associations between GO terms and taxa (Falda, et al. 2016). The taxonomic constraints can be visualized on a taxonomic tree at their website one GO term at a time. With the help of authors (Stefano Toppo, personal communication), we obtained a complete list of their constraints for all GO terms (A updated version compared with the results on the website). For each GO term, all taxa including both the leaves and the internal node are separated into three main categories: “in taxon” which means this GO term is allowed for the taxon, “never_in_taxon” which means the GO term is not allowed for the taxon, and “neutral” which they do not have any conclusions. The “in taxon” and “never_in_taxon” constraints for leaf taxon are summaries from gene annotations of species in the leaf taxa, while the constraints for internal nodes like “Eukaryota” are from propagations of the leaf taxa in the taxonomic tree.

We compared our taxon constraints with FunTaxIS results similarly to how we compared with annotations from experimental evidence, and found four types of conflicts: (1) "in taxon" taxon of FunTaxIS is outside of our “gained in” taxon, (2) "in taxon" taxon of FunTaxIS is within Loss taxon, (3) "never_in" taxon of FunTaxIS is within our Gain taxon and not within Loss taxon, (4) "never_in" taxon of FunTaxIS is an ancestor of our Gain taxon.

The first type of conflict is perhaps of most use for finding possible errors of our taxon constraints construction. FunTaxIS constrains taxa that are closer to the leaves of the tree of life than most of our constraints, thus this type of conflict is comparable with the type of conflict with experimental evidence that species with experimental evidences is outside of Gain taxon. There are 3313 conflict cases, covering 1266 GO terms. Manual examination shows many bizarre results, to name a few: The spindle pole body is the microtubule organizing center in yeast cells. According to FunTaxIS, term “spindle pole body separation” has “in_taxa” for Nematoda. “mitotic cell cycle” has “in_taxa” for “Cyanobcteria” and “Terrabacteria group”. A careful check of the GOA shows an experimental annotation to "negative regulation of spindle pole body separation" for a C.elegans gene. It is possible that FunTaxIS did not filter out the problematic GO annotations with experimental codes, and propagate the imprecise annotations to related GO terms. Another possibility is that the conflicts also root from imperfections of the experimental annotations. No examples of the second type of conflict are observed. We found a large number (32325 GO terms) of conflicts of the third type, where a “never_in” taxa of FunTaxIS is in our Gain taxon. It means that almost every GO term of our construction covers “legal” taxa which FunTaxIS considers “illegal” for gene annotations. For example, GO:000001 “mitochondrion inheritance”, our taxon constraint is “Eukaryota”, their “never_in” taxon include Viridiplantae, Bilateria, Vertebrata and Eukaryota. Our guess for potential reason of so many in-coherent “never_in” taxon is FunTaxIS uses current GO annotations to deduce the “never_in” and “in taxa” information. For terms that have rarely been annotated, there are high probabilities of it assigned with “never_in” constraints. This could also explain the fourth type of conflict which is a more extreme form the third type of conflict, “never_in” taxa of FunTaxIS is mom of Gain taxon instead of being within the Gain taxon. Example: “cellular bud site selection” is correctly constrained to “gain_at” Fungi in our taxon constraint construction, but the term is annotated with “never_in” Eukaryota based on FunTaxIS results.

## Future considerations

While we consider this initial version of taxon constraints expressed as evolutionary events to be far from complete, it is still a significant improvement over existing constraints. This version of taxon constraints is far from perfection. The manual taxon constraints in GO, PO, CL, FAO and UBERON may still contain errors and, for GO constraints particularly, an underestimate of constraints. For the time being, taxon constraints for ChEBI and PR terms are constructed from assembling experimental evidences, thus they are very likely to be constrained to in-conclusive taxon due to incompleteness of GO annotations. Our spot-checking of ChEBI-based constraints identified many errors; however, we couldn’t manually check constraints to thousands of GO terms. To minimize the errors, taxon constraints from these terms are only used when other sources of constraints are all missing.

There are about 11,000 GO terms to which we did not assign taxon constraints. However, most of them (61.9%) are molecular function terms. Because of the deep conservation of many molecular functions, we expect a relatively low coverage of taxon constraints. Even for the 4586 molecular function terms that have taxon constraints, the majority of them (90%) are constrained to root. Thus, we currently treat those with taxon constraints as “gain_at” LUCA, and allow for annotations to all cellular organisms. Besides, as some of the biological process terms without taxon constraints are linked to the molecular functions terms without constraints (250 out of 3231, 7.74%), these biological process GO terms are likely to also have constraints “gain_at” LUCA.

We are now sharing the taxon constraints with the broader GO Consortium, and we hope to continue to improve the constraints by adding further constraints manually in response to feedback from GO curators and the wider GO user community, and applying our automated propagation process. We also expect the ChEBI-based constraints to improve over time with the addition of more experimental annotations. However, our assumption of the most parsimonious evolutionary model, inheritance of the presence of a given endogenous chemical compound from a single common ancestor, is likely to be incorrect in many cases. We have applied the latest version of constraints to the Phylogenetic Annotation project of the GOC (Gaudet, et al. 2011). If curators try to annotate a gene from a species outside of the "gain_at" taxon, the system will warn the curators of potential conflicts. They will give us feedback if they find errors in our taxon constraints.

Note that the taxon constraints as currently constructed apply only to cellular organisms, and not to the functions of viral genes, as annotations to viral proteins are relatively rare and phylogenetic relationships are much more difficult to uncover. We plan to treat viral gene functions in the future.

The constraint file is freely available at https://github.com/haimingt/GOTaxonConstraint. We encourage users to give us feedback if they find errors.

## Conclusions

We have implemented a method of representing taxon constraints in terms of evolutionary gain and loss events relative to the species tree of life, and inferred constraints for the majority of GO by combining manual curation and rule based propagation. We first manually curate a seed list of high levels GO terms based on biological knowledge, making use of a broad range of ontologies that are linked to the GO and have not previously been applied to this problem. We propagate these constraints to related more specific GO terms based on the "true path rule" of the GO hierarchy. The constraints are then compared with experimental annotations to find conflicts. We manually check the conflicts and the GO terms without taxon constraints. Then we modify the seed list and begin another round of propagation. This process was repeated for several cycles of curation, propagation and correction.

By expressing taxon constraints in terms of evolutionary gains and losses, we can apply the existing knowledge about evolutionary histories, such as the evolution of eukaryotic cell, multicellularity and lineage-specific elaboration of anatomical structures, to the problem of constructing a computational representation of biological systems. We hope these constraints will provide valuable information for both GO curators and the wider community of GO and GO annotation users, and will prove useful in improving the accuracy of both manual and, in particular, computationally predicted, GO annotations.

## Acknowledgements

This work was supported by the National Science Foundation [grant number 1458808] and the National Human Genome Research Institute of the National Institutes of Health [award number U41HG002273].